\documentstyle[aps,prd]{revtex}
\begin{document}
\newcommand{\base}[2]{ {\scriptstyle (} #1_{\! \ssr{#2}} {\scriptstyle
)} }
\newcommand{\h}{\hat}
\newcommand{\ssr}[1]{\scriptscriptstyle{\rm \, #1}}
\thispagestyle{empty}
{\baselineskip0pt
\leftline{\large\baselineskip16pt\sl\vbox to0pt{\hbox{DAMTP}
               \hbox{University of Cambridge}\vss}}
\rightline{\large\today}
}
\vskip20mm
\begin{center}
{\Large\bf Gravity, Stability and Energy Conservation \\
 on the Randall-Sundrum Brane-World}
\end{center}

\begin{center}
{\large Misao Sasaki${}^{1,4}$,
Tetsuya Shiromizu${}^{2,4,5}$ and Kei-ichi Maeda${}^{3,6}$}
\vskip 3mm
\sl{${}^1$Department of Earth and Space Science, Graduate School of
Science,\\
Osaka University, Toyonaka 560-0043, Japan
\vskip 5mm
${}^2$DAMTP, University of Cambridge \\
Silver Street, Cambridge CB3 9EW, United Kingdom
\vskip 5mm
${}^3$Isaac Newton Institute, University of Cambridge, \\
20 Clarkson Road, Cambridge CB3 0EH, United Kingdom
\vskip 5mm
${}^4$Department of Physics, The University of Tokyo, Tokyo 113-0033, Japan
\vskip 5mm
${}^5$Research Centre for the Early Universe(RESCEU), \\
The University of Tokyo, Tokyo 113-0033, Japan
\vskip 5mm
${}^6$Department of Physics, Waseda University, Shinjuku,
Tokyo 169-8555, Japan}
\end{center}

\begin{abstract}
We carefully investigate the gravitational perturbation of the
Randall-Sundrum (RS) single brane-world solution [hep-th/9906064],
 based on a covariant curvature tensor formalism recently developed
by us. Using this curvature formalism, it is known that
the `electric' part of the 5-dimensional Weyl tensor, denoted by
$E_{\mu\nu}$, gives the leading order correction to
the conventional Einstein equations on the brane.
We consider the general solution of the perturbation equations
for the 5-dimensional Weyl tensor caused by the matter fluctuations
on the brane. By analyzing its asymptotic behaviour
in the direction of the 5th dimension, we find the
curvature invariant diverges as we approach the Cauchy horizon.
However, in the limit of asymptotic future in the vicinity of
 the Cauchy horizon, the curvature invariant falls off fast enough
to render the divergence harmless to the brane-world.
We also obtain the asymptotic behavior of $E_{\mu\nu}$ on the brane
at spatial infinity, assuming the matter perturbation is
localized. We find it falls off sufficiently fast and will not affect
the conserved quantities at spatial infinity. This
indicates strongly that the usual conservation law,
such as the ADM energy conservation, holds
on the brane as far as asymptotically flat spacetimes are
concerned.
\end{abstract}

\begin{center}
OUTAP-109;
DAMTP-1999-173;
UTAP-356;
RESCEU-49/99
\end{center}
\section{Introduction}

A recent discovery by Randall and Sundrum of the exact
solution that describes Minkowski branes in the 5-dimensional
anti-de Sitter space \cite{RS1,RS2} has attracted much attention.
In particular, in their second paper \cite{RS2},
they showed that a single Minkowski brane solution
is possible if the brane has positive tension, and
opened up the possibility of dimensional reduction without compactifying
extra dimensions.
Subsequently, a various aspects of, and variants of the
RS solution have been discussed by many authors.
Among them are work on non-linear plane-waves
\cite{SUGRA}, on black strings and black cigars
\cite{BH,Horowitz}, on the AdS/CFT correspondence \cite{CFT,CFT2},
on cosmological solutions
\cite{BDL2,muko,Flanagan,Kraus,Ida,Csaki1,Cline,Csaki2,Olive},
on the stability of the RS solution \cite{Tama} 
and on the quantum creation of the brane world\cite{Misao}(see also 
\cite{early,Cvetic})

Meanwhile, we formulated a covariant set of equations
that describes both the 5-dimensional gravity and the
4-dimensional gravity on the brane \cite{Tess}.
With $Z_2$-symmetry, which is expected to hold
from an $M$-theoretic point of view \cite{Witten,String},
we found the negative tension brane-world would not be allowed
since it would be a world of anti-gravity.
Then Garriga and Tanaka showed that the negative tension
brane is a world of a negative Brans-Dicke parameter 
at the linear perturbation order \cite{Tama}.

On the other hand, we found a positive tension brane has the correct
sign of gravity, and the equations reduce to the conventional
Einstein equations in the low energy limit, provided that
the extra term due to the `electric' part of the
5-dimensional Weyl tensor, $E_{\mu\nu}$, is negligible \cite{Tess}.
Thus it is urgently important to clarify the effect of
$E_{\mu\nu}$ to the brane-world.
Part of this program was first done by Randall and Sundrum themselves
\cite{RS2} and more rigourously by Garriga and Tanaka \cite{Tama},
and they showed that the effect is small at large distances
from the source. However, this conclusion
was obtained only for static perturbations. Furthermore, since they
adopted the metric perturbation formalism, the asymptotic behavior of
the curvature perturbations in 5 dimensions as the Cauchy horizon is
approached was not straightforward to see.

In this paper, we carefully investigate the first order perturbation of the
RS brane-world in terms of the 5-dimensional Weyl tensor.
We focus on the single brane model. 
We first briefly review our curvature tensor formalism. Then we
derive the evolution equation for $E_{\mu\nu}$, and give
an expression for the retarded Green function with appropriate boundary
condition on the brane. We recover the result obtained in
\cite{RS2,Tama} on the brane for static sources. For general
spacetime dependent sources, we evaluate the asymptotic behavior
of the Weyl tensor. We find the Weyl curvature invariant diverges
as we approach the Cauchy horizon. Nevertheless, an inspection of
the asymptotic behavior as we approach infinite future along near
the Cauchy horizon, this instability does not affect the brane-world
since it infinitely redshifts away. This infinite redshift effect
was suggested by Chamblin and Gibbons\cite{SUGRA}.
Then we discuss the energy conservation on the brane.
Local energy-momentum conservation is guaranteed by the
4-dimensional covariance.
We find globally conserved energy exists as well, just as
in the conventional Einstein gravity, for asymptotically flat spacetimes. 


\section{The effective Einstein equations on the brane}

In this section, we briefly review the effective gravitational
equations on the brane and the equations for the 5-dimensional
Weyl tensor which were derived in \cite{Tess}.

We consider a 5-dimensional spacetime with negative vacuum energy
but otherwise vacuum,
\begin{eqnarray}
 G_{\mu\nu}=\kappa_5^2T_{\mu\nu}\, ; \quad T_{\mu\nu}=-\Lambda g_{\mu\nu}\,,
\end{eqnarray}
and a brane in this spacetime as a fixed point of the $Z_2$-symmetry.
$\kappa_5$ is the 5-dimensional gravitational coupling
constant ($\kappa_5^2=8\pi G_5$).
We assume the metric of the form,
\begin{eqnarray}
 ds^2=\left(n_\mu n_\nu+q_{\mu\nu}\right)dx^\mu dx^\nu
=d\chi^2+q_{\mu\nu}dx^\mu dx^\nu\,,
\end{eqnarray}
where $n_\mu dx^\mu=d\chi$ is unit normal to the $\chi=$constant
hypersurfaces,
one of which corresponds to the brane, and $q_{\mu\nu}$ is the
induced metric on the $\chi=$constant hypersurfaces.
Then, thanks to the $Z_2$-symmetry, the effective 4-dimensional
gravitational equations on the brane take a form that resembles
the conventional Einstein equations:
%
\begin{eqnarray}
{}^{(4)}G_{\mu\nu}=-\Lambda_4q_{\mu\nu}
+ 8 \pi G_N\tau_{\mu\nu}+\kappa_5^4\,\pi_{\mu\nu}
-E_{\mu\nu}\,,
\label{eq:effective}
\end{eqnarray}
%
where
%
\begin{eqnarray}
\Lambda_4&=&\frac{1}{2}\kappa_5^2
\left(\Lambda +\frac{1}{6}\kappa_5^2\,\lambda^2\right)\,,
\label{Lamda4}\\
G_N&=&{\kappa_5^4\,\lambda\over48 \pi}\,,
\label{GNdef}\\
\pi_{\mu\nu}&=&
-\frac{1}{4} \tau_{\mu\alpha}\tau_\nu^{~\alpha}
+\frac{1}{12}\tau\tau_{\mu\nu}
+\frac{1}{8}q_{\mu\nu}\tau_{\alpha\beta}\tau^{\alpha\beta}-\frac{1}{24}
q_{\mu\nu}\tau^2\,,
\label{pidef}\\
E_{\mu\nu}& = & C_{\mu\alpha\nu\beta}n^\alpha n^\beta
\end{eqnarray}
%
and $C_{\mu\nu\alpha\beta}$ is the 5-dimensional Weyl tensor.
$\lambda$ is the vacuum energy on the brane and gives the brane tension.
$\tau_{\mu\nu}$ is the energy-momentum tensor of the matter on
the brane. It should be noted that $E_{\mu\nu}$ here is not the one
exactly on the brane (which is proportional to the delta function), but
the one that is evaluated by taking the limiting value to the brane.
But for simplicity, we call it $E_{\mu\nu}$ on the brane in the rest of
the paper.
It can be evaluated from either side of the brane due to
$Z_2$-symmetry.

As $G_N$, which corresponds to the 4-dimensional Newton constant, has
the same sign as $\lambda$, we assume a positive tension brane to obtain
the conventional gravity on the brane. In the case of a two brane model with
positive and
negative tensions, it has been argued that it is possible to recover
the normal gravity if the distance between the two branes is fine-tuned
\cite{Tama} and if some mechanism to stabilize the distance is introduced
\cite{Csaki2}. In this paper, however, we focus on a single brane model
and assume $\lambda>0$.

The difference of Eq.~(\ref{eq:effective}) from the Einstein gravity
is the presence of $\pi_{\mu\nu}$ and $E_{\mu\nu}$ on the right-hand side.
With $\kappa_5$ and $\lambda$ of a very high energy scale, it is
easy to see that $\pi_{\mu\nu}$ can be safely neglected in the low
energy limit. On the other hand, since $E_{\mu\nu}$ comes from the
5-dimensional Weyl tensor, there is no {\it a priori}\/ reason to expect
that it is small even in the low energy limit.
In fact, in the two-brane model, it is this part that contributes
dominantly to recover Einstein gravity on the negative tension
brane when the distance between the branes is fine-tuned \cite{Tama}.
We note that because of the contracted Bianchi identities, we have
\begin{eqnarray}
 D^\mu E_{\mu\nu}=\kappa_5^4 D^\mu\pi_{\mu\nu},
\label{divE}
\end{eqnarray}
where $D_\mu$ is the covariant differentiation with respect to
$q_{\mu\nu}$.
Hence, in the low energy limit when we can neglect $\pi_{\mu\nu}$,
$E_{\mu\nu}$ is transverse-traceless with respect to $q_{\mu\nu}$.

The effective gravitational equations on the brane (\ref{eq:effective})
are not closed but one must solve the gravitational field in the bulk at
the same time. The 5-dimensional equations in the bulk we have to solve are
%
\begin{eqnarray}
\mbox \pounds_n E_{\alpha\beta} = &&
D^\mu B_{\mu(\alpha\beta)}
+ \frac{1}{6}\kappa_5^2
\Lambda\left(K_{\alpha\beta}-q_{\alpha\beta}K\right)
+K^{\mu\nu}{}^{(4)}R_{\mu\alpha\nu\beta}
 \nonumber \\
&& +3K^\mu{}_{(\alpha}E_{\beta)\mu}-KE_{\alpha\beta}
+\left(K_{\alpha\mu}K_{\beta\nu}
-K_{\alpha\beta}K_{\mu\nu}\right)K^{\mu\nu}\,,
\label{eq:EEE}
\\
\mbox \pounds_n {B}_{\mu\nu\alpha}=&&-
2D_{[\mu}{E}_{\nu]\alpha}+K_\alpha{}^\sigma
{B}_{\mu\nu\sigma}
-2{B}_{\alpha \sigma [\mu }K_{\nu]}{}^\sigma\,,
\label{eq:BBB}
\\
\mbox \pounds_n {}^{(4)}R_{\mu \nu \alpha \beta}
=&&-2{}^{(4)}R_{\mu\nu\sigma [\alpha}K^\sigma_{\beta]}
-2D_{[\mu}{B}_{|\alpha\beta|\nu]},
\label{eq:bianchi3}
\end{eqnarray}
%
where $K_{\mu\nu}=(1/2)\mbox \pounds_n q_{\mu\nu}$ and
$B_{\mu\nu\alpha}= q_\mu^{~\rho} q_\nu^{~\sigma}
C_{\rho\sigma\alpha\beta}n^\beta$.
These equations are derived from the 5-dimensional Bianchi identities
\cite{Tess}. 

The equation for $E_{\mu\nu}$ has an alternative form that
will be more convenient for later use;
%
\begin{eqnarray}
\mbox \pounds_n E_{\alpha\beta}  =&&
D^\mu B_{\mu(\alpha\beta)}
+ K^{\mu\nu}{}^{(4)}C_{\mu\alpha\nu\beta}
+4K^\mu_{~(\alpha}E_{\beta)\mu}
-{3\over2}KE_{\alpha\beta}
-{1\over2}q_{\alpha\beta}K^{\mu\nu}E_{\mu\nu}
\nonumber\\
&&
+2{\tilde K}^\mu_{~\alpha}{\tilde K}_{\mu\nu}{\tilde K}^{\nu}_{~\beta}
-\frac{7}{6}{\tilde K}_{\mu\nu}{\tilde K}^{\mu\nu}{\tilde
K}_{\alpha\beta}\,,
\label{eq:alter}
\end{eqnarray}
%
where ${\tilde K}_{\alpha\beta}$ is the traceless part of
$K_{\alpha\beta}$,
%
\begin{eqnarray}
{\tilde K}_{\alpha\beta}=K_{\alpha\beta}-\frac{1}{4}q_{\alpha\beta}
K^{\mu}_{~\mu}\,.
\end{eqnarray}
%

Equations (\ref{eq:EEE}) (or (\ref{eq:alter})),
 (\ref{eq:BBB}) and (\ref{eq:bianchi3}) are to be solved under
the boundary condition at the brane,
\begin{eqnarray}
& & D^\mu E_{\mu\nu}|_{\rm brane}= \kappa_5^4 D^\mu\pi_{\mu\nu}\\
& & {B}_{\mu\nu \alpha}|_{\rm
brane}=2D_{[\mu}K_{\nu]\alpha}|_{\rm brane}=
-\kappa_5^2D_{[\mu}\Bigl( \tau_{\nu] \alpha}
-\frac{1}{3}q_{\nu] \alpha}\tau \Bigr),
\label{bcwall}
\end{eqnarray}
where we used the expression of $K_{\mu\nu}$ which
is obtained by the Israel junction condition and
$Z_2-$symmetry,
\begin{eqnarray}
K_{\mu\nu}|_{\rm brane}=-\frac{1}{6}\kappa_5^2 \lambda q_{\mu\nu}
-\frac{1}{2}\kappa_5^2
\Bigl( \tau_{\mu\nu} -\frac{1}{3}q_{\mu\nu}\tau\Bigr).
\end{eqnarray}

\section{The First order Perturbation around the RS-brane world}
In this section, we derive the equations for the first order perturbation
around the RS solution\cite{RS2}. As we have noted in the
previous section, we focus on the case of a single brane with positive
tension.

\subsection{The perturbation equations}

The background bulk spacetime is taken to be
the anti-de Sitter spacetime whose metric is
%
\begin{eqnarray}
ds^2=d\chi^2+e^{-2\chi/\ell}\eta_{ij}dx^idx^j
=\frac{\ell^2}{z^2}\Bigl[dz^2+ \frac{1}{\ell^2}\eta_{ij}dx^i dx^j\Bigr],
\end{eqnarray}
%
where $\ell={\sqrt {-6/\Lambda\kappa_5^2}}$,
$z:=e^{\chi/\ell}$, $i=0$, $1$, $2$, $3$ and $\eta_{ij}$ is the
metric of the Minkowski spacetime, $\eta_{ij}dx^i dx^j
=-dt^2+d\bbox{x}^2$. $\chi=\infty$ or $z=\infty$
is the Cauchy horizon. The RS solution is obtained by
putting a brane on a surface $z=z_*$
and glueing two identical copies of the region $z \geq z_*$ of the AdS
\cite{RS1,RS2}. Since the coordinate $z$ is scale-free, we may set
$z_*=1$ without loss of generality. This solution is obtained
from Eq.~(\ref{eq:effective}) by putting
$\Lambda_4=0$ and $\tau_{\mu\nu}=E_{\mu\nu}=0$.

We now consider the first order perturbation of the RS solution.
We assume $\tau_{\mu\nu}$ is a small quantity of $(\epsilon)$
and solve the perturbation linear in $\epsilon$.
Hence, we have
%
\begin{eqnarray}
{}^{(4)}G_{\mu\nu}= 8 \pi G_N \tau_{\mu\nu}-E_{\mu\nu}+O(\epsilon^2)
\end{eqnarray}
%
on the brane.

In the first order of $\tau_{\mu\nu}$, the extrinsic curvature
$K_{\mu\nu}$ can be written as
%
\begin{eqnarray}
K_{\alpha\beta}=-\frac{1}{\ell}q_{\alpha\beta}+k_{\alpha\beta},
\label{Kpert}
\end{eqnarray}
%
where $k_{\alpha\beta}$
is $O(\epsilon)$. The Lie derivative of $k_{\mu\nu}$ with respect to $n^\mu$
is just $E_{\mu\nu}$ apart from the signature,
%
\begin{eqnarray}
\mbox \pounds_n k_{\alpha\beta}=k_\alpha{}^{\mu}k_{\beta\mu}
-E_{\alpha\beta} =-E_{\alpha\beta}+O(\epsilon^2).
\label{kLie}
\end{eqnarray}
%

Using Eq.~(\ref{Kpert}), Eq.~(\ref{eq:alter}) simplifies to
%
\begin{eqnarray}
\mbox \pounds_n E_{\alpha\beta}=D^\mu B_{\mu(\alpha\beta)}
+\frac{2}{\ell}E_{\alpha\beta}+O(\epsilon^2).
\label{Elin}
\end{eqnarray}
%
On the other hand, Eq.~(\ref{eq:BBB}) reduces to
\begin{eqnarray}
\mbox \pounds_n B_{\mu\nu\alpha}=-2D_{[\mu}E_{\nu] \alpha}+O(\epsilon^2).
\label{Blin}
\end{eqnarray}
Combining these two equations, we obtain the wave equation for
$E_{\mu\nu}$ in 5 dimensions,
%
\begin{eqnarray}
\Bigl[ \mbox \pounds_n \Bigl(   \mbox \pounds_n
+K_0 \Bigr) +D^2 +\frac{4}{\ell^2} \Bigr]E_{\alpha\beta}
=\Bigl[ \Box_5 +\frac{4}{\ell^2} \Bigr]E_{\alpha\beta}=0,
\label{eq:green}
\end{eqnarray}
%
where $\Box_5$ is the 5-dimensional d'Alembertian.
In the coordinates $(z,x^\mu)$, this becomes
%
\begin{eqnarray}
\Bigl[\partial_z^2-\frac{3}{z}\partial_z +\ell^2 \Box_4+\frac{4}{z^2}
\Bigr]E_{\alpha\beta}=0,
\end{eqnarray}
%
where $\Box_4$ is the 4-dimensional d'Alembertian.
Substituting Eq.~(\ref{bcwall}) to Eq.~(\ref{Elin}),
we obtain the boundary condition of $E_{\mu\nu}$ on the brane,
%
\begin{eqnarray}
\partial_z \Bigl ( \frac{E_{\mu\nu}}{z^2}\Bigr)_{z=1}
=-{1\over2}\kappa_5^2\ell
\Bigl[D^2 \tau_{\mu\nu}+\frac{1}{3}D_\mu D_\nu \tau
-\frac{1}{3}q_{\mu\nu} D^2 \tau \Bigr].
\label{eq:EBC}
\end{eqnarray}
%
In addition to the above, since we have $D^\mu E_{\mu\nu}=O(\epsilon^2)$
on the brane, and $\mbox\pounds_n D^\mu E_{\mu\nu}=O(\epsilon^2)$,
$E_{\mu\nu}$ is transverse-traceless with respect to the 4-metric
$q_{\mu\nu}$ everywhere in the bulk.


Once $E_{\mu\nu}$ is solved, we can integrate Eqs.~(\ref{eq:BBB})
and (\ref{eq:bianchi3}) to obtain $B_{\alpha\beta\mu}$
and ${}^{(4)}C_{\alpha\beta\mu\nu}$.
To the first order, these are expressed in the coordinates $(z,x^i)$ as
%
\begin{eqnarray}
\partial_z B_{\mu\nu\alpha}=-\frac{2\ell}{z}\partial_{[\mu}E_{\nu]
\alpha}
\label{eq:1difB}
\end{eqnarray}
\begin{eqnarray}
\partial_z (z^2 {}^{(4)}C_{\mu\nu\alpha\beta})
=\ell z [\partial^\rho B_{\rho (\mu [\alpha)}\eta_{\beta]\nu}
-\partial^\rho B_{\rho (\nu[\alpha)} \eta_{\beta] \mu} ]-2 \ell z
\partial_{[\mu}B_{|\alpha \beta| \nu]}
\label{eq:1difC}
\end{eqnarray}
The above equations will be used when we evaluate the asymptotic behaviour
of the Weyl curvature near the Cauchy horizon.
To discuss the behaviour near the Cauchy horizon, we consider an
invariant quantity defined by the 5-dimensional Weyl tensor,
%
\begin{eqnarray}
C_{\mu\nu\alpha\beta}C^{\mu\nu\alpha\beta}
={}^{(4)}C_{\mu\nu\alpha\beta}{}^{(4)}C^{\mu\nu\alpha\beta}
+6 E_{\mu\nu}E^{\mu\nu}+4 B_{\mu\nu\alpha}B^{\mu\nu\alpha}+O(\epsilon^3).
\label{Winv}
\end{eqnarray}
%

\subsection{The retarded Green function}

The general solution of $E_{\mu\nu}$ can be expressed in terms of the
Green function satisfying
%
\begin{eqnarray}
(\Box_5+4/\ell^2)G(z,x)=-\delta^5 (z,x).
\end{eqnarray}
%
As usual, we assume there is no incoming waves from past Cauchy horizon.
Hence we consider the retarded Green function. In addition to this
causal boundary condition, we have one more boundary condition at the
brane.  Because of the boundary condition for $E_{\mu\nu}$,
Eq.~(\ref{eq:EBC}), we must impose the corresponding condition on the
Green function,
\begin{eqnarray}
\partial_z \Bigl( \frac{G}{z^2}\Bigr)_{z=1}=0.
\label{eq:GBC}
\end{eqnarray}
With the Green function satisfying this condition,
$E_{\mu\nu}$ is given by
%
\begin{eqnarray}
E_{\mu\nu}(z,x) &=& \int_{z'=1} d^4x' (-g(x'))^{1/2}g^{zz}(x')
\Bigl[\partial_{z'}G(z,x;\,z',x')E_{\mu\nu}(x')
   -G(z,x;\,z',x') \partial_{z'}E_{\mu\nu}(x')\Bigr]
\nonumber\\
&=&-\int_{z'=1} d^4x' (-g(x'))^{1/2}g^{zz}(x')
G(z,x;\,z',x') \partial_{z'}\left({E_{\mu\nu}(x')\over z'^2}\right)
\nonumber\\
&=&{\kappa_5^2\over2}
 \int_{z'=1} d^4 x' G(z,x;\,z',x')
\Bigl( \Box_4 \tau_{\mu\nu}+\frac{1}{3}\partial_\mu \partial_\nu \tau
-\frac{1}{3}q_{\mu\nu}\Box_4 \tau(x')\Bigr).
\label{eq:EF}
\end{eqnarray}
%
For bounded sources, which we are mainly interested in,
the last line of the above equation can be re-expressed as
\begin{eqnarray}
 E_{\mu\nu}(z,x)={\kappa_5^2\over 2}
\left[\delta_\mu^\alpha \delta_\nu^\beta\Box_4
+{1\over3}\eta^{\alpha\beta}\left(\partial_\mu\partial_\nu
-\eta_{\mu\nu}\Box_4\right)\right]_x \int_{z'=1} d^4 x'
G(z,x;\,z',x')\tau_{\alpha\beta}(x').
\label{Etau}
\end{eqnarray}

Now let us construct the retarded Green function.
The general form of a mode function satisfying Eq.~(\ref{eq:green}) is
%
\begin{eqnarray}
u_{m,p}(z,x)=N_m z^2 \left(J_0(mz)+b_m N_0(mz)\right)
\frac{e^{-i\omega_{pm} t+i\bbox{p}\cdot\bbox{x}}}
{(2\pi)^{3/2}{\sqrt {2 \omega_{pm}}} },
\end{eqnarray}
%
where $\omega_{pm}=\sqrt{\bbox{p}^2+m^2/\ell^2}$ and
$N_m$ is a normalization constant. The boundary condition (\ref{eq:EBC})
requires that the mode functions should satisfy (\ref{eq:GBC}),
which determines $b_m$ as
\begin{eqnarray}
b_m=-\frac{J_1(m)}{N_1(m)}\,.
\end{eqnarray}
Then $N_m$ is fixed by requiring that the mode functions should be
properly normalized with respect to the Klein-Gordon norm.
This gives
\begin{eqnarray}
N_m = \frac{m^{1/2}}{{\sqrt {(1+b_m^2)\ell}}}\,.
\end{eqnarray}
Using the mode functions obtained above,
the retarded Green function is constructed as
%
\begin{eqnarray}
G_{\rm R}(z,x;\,z',x')|_{z'=1}
 =  -\frac{2z^2}{\pi \ell}\int_0^\infty dm
\frac{N_1(m)J_0(mz)-J_1(m)N_0(mz)}{N_1(m)^2+J_1(m)^2}g_m(x,x')
\label{Gform}
\end{eqnarray}
%
where $g_m(x,x')$ is the 4-dimensional Green function given by
%
\begin{eqnarray}
g_m(x,x')=\int \frac{d\omega d^3p}{(2\pi)^4}
\frac{e^{-i\omega(t-t')+i\bbox{p} \cdot (\bbox{x}-\bbox{x}')}}
{(m/\ell)^2+\bbox{p}^2-(\omega + i \epsilon)^2}\,.
\end{eqnarray}
%

\subsection{Static case}

First let us consider the static case. In this case, the
Green function is given by the integration of the retarded
Green function over the time. Then the 4-dimensional part
gives rise to the factor $e^{-mr/\ell}$. Hence the integration with
respect to $m$ is dominated by the contribution of small $m$, and
we obtain
%
\begin{eqnarray}
G(z,\bbox{x};1,\bbox{x}')=\int^\infty_{-\infty} dt\, G_{\rm R}
(z,t,\bbox{x};1,0,\bbox{x}')
\simeq \frac{\ell z^2}{4\pi [r^2+(\ell z)^2]^{3/2}},
\end{eqnarray}
%
where $r=|\bbox{x}-\bbox{x}'|$.

On the brane the far field approximation gives
%
\begin{eqnarray}
G(1,\bbox{x};1, \bbox{x}') \simeq \frac{\ell}{4\pi r^3}.
\end{eqnarray}
%
{}From Eq. (\ref{Etau}), we obtain the leading order of $E_{\mu\nu}$ as
%
\begin{eqnarray}
&&E_{00} \simeq
\frac{4G_N \ell^2}{r^5}\int d^3 x \Bigl(\rho+\frac{3}{2}P\Bigr) ,
\nonumber\\
&&E_{ij} \simeq {1\over3}\delta_{ij}E_{00}\,,\qquad E_{0i} \simeq 0,
\label{eq:statice00}
\end{eqnarray}
%
where we assumed the perfect fluid form, $\tau_{\mu\nu}=\rho u_\mu
u_\nu +P (q_{\mu\nu}+u_\mu u_\nu)$, for the energy momentum tensor.
The result is consistent with those obtained in Refs. \cite{RS2,Tama}.

For $z \gg 1$, that is, near the Cauchy horizon, the static
Green function is approximately given by
%
\begin{eqnarray}
G(z,\bbox{x};1, \bbox{x}')\simeq \frac{1}{4\pi\ell^2 z }
\Bigl[1-\frac{3}{2}\Bigl(\frac{r}{\ell z} \Bigr)^2  \Bigr].
\end{eqnarray}
%
Hence we obtain the asymptotic behaviour as
%
\begin{eqnarray}
E_{\mu\nu} =O(\partial_r^2G)=O(z^{-3}),
\quad\partial_\mu E_{\alpha\beta}=O(z^{-5})\,.
\end{eqnarray}
%
%
With the help of Eqs.~(\ref{eq:1difB}) and (\ref{eq:1difC}),
the above gives
\begin{eqnarray}
B_{\mu\nu\alpha} =O(z^{-5}),\quad
{}^{(4)}C_{\mu\nu\alpha\beta}=O(z^{-7}).
\end{eqnarray}
%
Therefore we have
%
\begin{eqnarray}
E_{\mu\nu}E^{\mu\nu} =O(z^{-2}),\quad
B_{\mu\nu\alpha}B^{\mu\nu\alpha}=O(z^{-4}),
\quad
{}^{(4)}C_{\mu\nu\alpha\beta}{}^{(4)}C^{\mu\nu\alpha\beta}=O(z^{-6}),
\end{eqnarray}
%
from which we obtain the estimate,
%
\begin{eqnarray}
C_{\mu\nu\alpha\beta} C^{\mu\nu\alpha\beta}=O(z^{-2}).
\end{eqnarray}
%
This means that the perturbation remains regular at the
Cauchy horizon for the static case.

\subsection{General case}

We now turn to the general case of time-dependent sources.
We first note that the 4-dimensional Green function, $g_m(x,x')$,
can be evaluated exactly as
%
\begin{eqnarray}
g_m(x,x') =  \frac{1}{2\pi} \theta (t-t') \frac{\partial}{\partial s^2}
\Bigl[ \theta (s^2)J_0\Bigl(m\frac{|s|}{\ell} \Bigr) \Bigr]
\end{eqnarray}
%
where $s^2=(t-t')^2-r^2$ and $r=|\bbox{x}-\bbox{x}'|$.
Inserting this expression to Eq.~(\ref{Gform}), we find
the part that contains the derivative of $\theta(s^2)$
(hence is proportional to $\delta(s^2)$) will not contribute.
This is because this part of $g_m$ becomes independent of $m$,
hence the integration with respect to $m$ gives $\delta(z-1)$, but
$z=1$ is outside the domain of definition of $E_{\mu\nu}$ (recall that
$E_{\mu\nu}$ on the brane is actually defined by taking the limit
to the brane). Thus we have
\begin{eqnarray}
G_{\rm R}(z,x;\,z',x')|_{z'=1}
 =  -\frac{z^2}{\pi^2 \ell}\theta(t-t')\theta(s^2)
{\partial\over\partial s^2}\int_0^\infty dm
\frac{N_1(m)J_0(mz)-J_1(m)N_0(mz)}{N_1(m)^2+J_1(m)^2}
J_0\Bigl(m \frac{|s|}{\ell} \Bigr).
\end{eqnarray}

To evaluate the asymptotic behavior of $G_{\rm R}$, we may approximate the
above integral by assuming $m\gg1$. Physically this is because only the
high frequency modes (in the 5-dimensional sense) can reach the
null infinity, which corresponds to the Cauchy horizon in the present
case. Since $z>1$ we can then use the asymptotic form of the Bessel
functions except the function $J_0(m|s|)$; we keep it as it is since we
do not want to restrict the range of $s^2$.
Thus, we obtain
%
\begin{eqnarray}
G_{\rm R}(z,x;\,z',x')|_{z'=1}
&\simeq&
-{\theta(t-t')\theta(s^2)z^{3/2}\over\pi^2\ell}{\partial\over\partial s^2}
\int_0^\infty dm\cos [m(z-1)]J_0\Bigl(m\frac{|s|}{\ell}\Bigr)
\nonumber\\
&=&\theta (t-t')\theta (s^2)\theta (s^2-\ell^2(z-1)^2)\,
\frac{z^{3/2}}{2\pi^2[s^2-\ell^2(z-1)^2]^{3/2}}\,.
\end{eqnarray}
%

On the brane ($z=1+0$), the Green function becomes
%
\begin{eqnarray}
G_{\rm R}(1,x;1,x') \simeq
\frac{\theta (s^2)\theta (t-t')}{2\pi^2 s^3}\,,
\end{eqnarray}
%
which is in accordance with a naive expectation.
This implies the behaviour of $E_{\mu\nu}$ in the far-field region as
%
\begin{eqnarray}
E_{\mu\nu}\sim O(\partial_s^2 G_R)=O(s^{-5}).
\end{eqnarray}
%
This should be compared with the energy momentum tensor of a radiative
field, $\tau_{\mu\nu}\sim s^{-2}$ at null infinity.
Thus we conclude that $E_{\mu\nu}$
cannot carry away the energy momentum from a system to infinity.
We will come back to this point in the next section.

To investigate the asymptotic behavior near the Cauchy horizon,
it is more convenient to work in the null coordinates,
$u=s-\ell (z-1)$ and $v=s+\ell (z+1)$. Then with the help
of Eqs.~(\ref{eq:1difB}) and (\ref{eq:1difC}), we find
%
\begin{eqnarray}
&&E_{\mu\nu}=O(\partial_s^2 G_R)=O\Bigl(\frac{(v-u)^{3/2}}
{(vu)^{5/2}}  \Bigr),
\nonumber\\
&&B_{\mu\nu\alpha}=O(\partial_\alpha E_{\mu\nu})
=O(\partial_s^3 G_R)=O\Bigl(\frac{(v+u)(v-u)^{3/2}}{(vu)^{7/2}}  \Bigr),
\nonumber\\
&&{}^{(4)}C_{\mu\nu\alpha\beta}=O(\partial_\alpha B_{\mu\nu\beta})
=O(\partial_s^4 G_R)=O\Bigl(\frac{(v-u)^{3/2}}{(vu)^{7/2}}  \Bigr).
\end{eqnarray}
%
We take the limit $v\to\infty$ along $u=$constant to approach the
Cauchy horizon. Then
$E_{\mu\nu}=O(v^{-1})$, $B_{\mu\nu\alpha}=O(v^{-1})$ and
${}^{(4)}C_{\mu\nu\alpha\beta}=O(v^{-2})$.
Therefore, the curvature invariant~(\ref{Winv})
diverges in this limit as
\begin{eqnarray}
 C_{\mu\nu\alpha\beta}C^{\mu\nu\alpha\beta}=O(v^{4}).
\end{eqnarray}
This means that the Cauchy horizon is unstable to the perturbation.
The bad behaviour of the Weyl tensor, however, does not necessarily
imply an instability of the brane-world. It will not affect the
brane if the divergence disappears in the limit of distant future;
$u=\alpha v\to\infty$ ($\alpha=$const.$<1$).
This is the infinite redshift effect. In the present case,
if we take this limit,
we find $C_{\mu\nu\alpha\beta}C^{\mu\nu\alpha\beta}=O(u^{-3})$.
Hence the brane-world is unaffected by the instability of the Cauchy
horizon.

\section{The energy conservation on the brane}

In the conventional Einstein gravity,
the ADM energy is conserved if the energy-momentum tensor
$\tau_{\mu\nu}$decays faster than $\sim r^{-3}$ towards spatial infinity.
For the matter source of compact support, this condition is
trivially satisfied \cite{AH,AR,Gen}.
In the present case, we have $E_{\mu\nu}$ which is a part
of the 5-dimensional Weyl curvature, in addition to the matter
energy-momentum tensor. However, as we have seen in the previous
section, although the 4-geometry on the brane is affected by
$E_{\mu\nu}$, the effect seems to be quite subtle:
 First, $E_{\mu\nu}$ itself is locally conserved
at the linear order, independent of $\tau_{\mu\nu}$.
Second, both the static and dynamic perturbations fall off
sufficiently rapidly at large distances from the source.
So we expect that the conservation of the total energy
holds also in the brane-world, provided the brane geometry
is asymptotically flat.

To re-confirm this expectation, in this section we present a more
detailed discussion of the energy conservation.
A brief review of the asymptotic structure at spatial infinity is given
in Appendix A. The following argument is based on a recently developed
formalism of the conformal infinity \cite{AR,Gen}, which is
much easier to deal with than the old formalism \cite{AH}.

First we express the asymptotic behavior of the 4-dimensional
Ricci tensor as
\begin{eqnarray}
L_{\mu\nu}={}^{(4)}R_{\mu\nu}-\frac{1}{6}q_{\mu\nu}{}^{(4)}R
\simeq L_{\mu\nu}^{(0)} + L_{\mu\nu}^{(1)} r^{-1}
+ L_{\mu\nu}^{(2)} r^{-2}+  L_{\mu\nu}^{(3)} r^{-3}
+O\Bigl( \frac{1}{r^4}\Bigr).
\end{eqnarray}
The total energy of a system is naturally defined by the electric part
of the 4-dimensional Weyl tensor,
${}^{(4)}E_{\mu\nu}={}^{(4)}C_{\mu\alpha\nu\beta}r^\alpha r^\beta$,
because it is the one that describes
the tidal force. $r^\alpha$ is the unit vector of the radial
direction. By a conformal transformation, one can expand the
spatial infinity to introduce the structure of a unit 3-dimensional
timelike hyperboloid. We denote the metric and the covariant derivative
of it by $p_{\mu\nu}$ and ${\cal D}_\mu$, respectively.
Then, assuming $L_{\mu\nu}^{(0)}=L_{\mu\nu}^{(1)}=L_{\mu\nu}^{(2)}=0$,
the leading order behavior of ${}^{(4)}E_{\mu\nu}$ for
$r\to\infty$ can be expressed as
%
\begin{eqnarray}
{}^{(4)}E_{\mu\nu} \simeq 
\frac{1}{r}\Bigl[ 
\frac{1}{2}({\cal D}_\mu {\cal D}_{\nu}
+p_{\mu\nu}) F^{(1)}-\frac{1}{2}(L_{\mu\nu}^{(3)}+p_{\mu\nu}
L_{\alpha\beta}^{(3)}r^\alpha r^\beta) \Bigr]+O\Bigl(
\frac{1}{r^2}\Bigr),
\label{eq:4dimE}
\end{eqnarray}
where $F^{(1)}$ is a function on the unit 3-dimensional timelike 
hyperboloid. In the present case, we have contributions from not only
$\tau_{\mu\nu}$ but also $E_{\mu\nu}$ to the Ricci tensor. 
{}From the analysis in the previous section at the spatial 
infinity,
$E_{\mu\nu}=O(r^{-5})$.
Hence, if the matter source is localized, we have
$L_{\mu\nu}^{(n)}=0$ for $n=0$, $1$, $2$, $3$.
Thus,
%
\begin{eqnarray}
{}^{(4)}E_{\mu\nu} \simeq
\frac{1}{2r}({\cal D}_\mu {\cal D}_{\nu}
+p_{\mu\nu}) F^{(1)}   + O\Bigl( \frac{1}{r^2}\Bigr),
\end{eqnarray}
%
Since ${}^{(4)}E_{\mu\nu}$ is traceless, ${}^{(1)}F$ satisfies
%
\begin{eqnarray}
({\cal D}^2+3)F^{(1)}=0.
\end{eqnarray}
%
Using this and introducing
${}^{(4)}{\tilde E}_{\mu\nu}=r\,{}^{(4)}E_{\mu\nu}$, we can show
%
\begin{eqnarray}
{\cal D}^\mu {}^{(4)}{\tilde E}_{\mu\nu}=0.
 \label{eq:div}
\end{eqnarray}
%
Now we define the ADM energy by
%
\begin{eqnarray}
E_{\rm ADM}:=-\frac{1}{8\pi} \int_{\cal C}
dS^{\mu\nu} \epsilon_{\mu\nu\alpha}{\tilde E}^{\alpha\beta}t_\beta
\label{eq:adm}
\end{eqnarray}
%
where where $t^\mu$ is the asymptotic conformal Killing
vector associated with the global time-translation symmetry induced
at spatial infinity and
${\cal C}$ is a 2-dimensional closed surface in the 3-dimensional
unit timelike hyperboloid. The above integration is invariant under
the change of the closed surface ${\cal C}$ because Eq.~(\ref{eq:div})
gives ${\cal D}_\mu ({}^{(4)}{\tilde E}^{\mu\nu}t_{\nu})=0$. This means
that $E_{\rm ADM}$ is conserved.

\section{Summary and Discussion}

We have carefully investigated the asymptotic behavior of the
linear perturbation around the single brane solution of the RS
brane-world scenario \cite{RS2}.
It is known that the effect of the
5-dimensional curvature appears in the form of $E_{\mu\nu}$, which is
the `electric' part of the 5-dimensional Weyl tensor, on the right-hand
side of the effective Einstein equations on the brane \cite{Tess}.
To deal with this part of the Weyl tensor directly, we have
employed the curvature tensor perturbation formalism developed in
\cite{Tess} instead of the metric perturbation formalism.

For static perturbations, we have found that $E_{\mu\nu}$
on the brane falls off as $r^{-5}$ from the source and
the Weyl curvature in the bulk behaves regularly near the Cauchy horizon
($z=\infty$) as $C_{\mu\nu\alpha\beta}C^{\mu\nu\alpha\beta}=O(z^{-2})$,
 in accordance with the previous result obtained in the metric
 perturbation formalism \cite{RS2,Tama}.

 For generic perturbations, we have found the perturbation
diverges as $C_{\mu\nu\alpha\beta}C^{\mu\nu\alpha\beta}=O(v^4)$ near the
Cauchy horizon ($v=\infty$, $u=$const.). This implies instability
of the Cauchy horizon. However, we have found that this instability
does not affect the brane-world because
$C_{\mu\nu\alpha\beta}C^{\mu\nu\alpha\beta}=O(u^{-3})$ for
$u=\alpha v\to\infty$ ($\alpha<1$) due to the infinite redshift effect.
This supports the conjecture by Chamblin and Gibbons \cite{SUGRA}
that the Cauchy horizon instability would not affect the brane-world.

We have also found that $E_{\mu\nu}$ decays
rapidly as $s^{-5}$ on the brane for generic spacetime-dependent
perturbations, where $s^2=t^2-r^2$. This is to be contrasted with
the case of a radiative field (like electromagnetic radiation)
that behaves like $s^{-2}$(To avoid the confusion,
a radiation field behaves like $\sim r^{-4}$ at spatial infinity.).
This implies $E_{\mu\nu}$ does not
carry energy away from the system to infinity.
Together with the fact that $E_{\mu\nu}=O(r^{-5})$ for static
perturbations, we have concluded that the ADM energy is well-defined
also in the brane-world and it is conserved.

One remaining issue is the positivity of the ADM energy. In the
linear perturbation, the right-hand side of the effective Einstein
equations has the additional term, $-E_{\mu\nu}$.
If $-E_{\mu\nu}u^\mu u^\nu>0$, where $u^\mu$ is an arbitrary
timelike vector, the brane will be stable in the semi-classical
level. In this connection, recent work on cosmological solutions
that $-E_{00}$ is proportional to the mass of the
5-dimensional Schwarzshild-AdS spacetime \cite{Flanagan,Kraus,Ida,Misao}
implies it is positive.
On the contrary, our result of the perturbation
analysis given in Eq.~(\ref{eq:statice00}) has
the opposite sign.
One might suspect that our result is incorrect.
However, there is a good reason to believe that
our result is indeed correct. Remember that $E_{00}$
is the tidal force in the 5th direction.
When there is a mass in the bulk,
this means we have a force that tends to `tear apart'
the matter on the brane in the 5th direction,
hence gives a repulsive contribution. This means
the energy is positive, in accordance with
cosmological results obtained in \cite{Flanagan,Kraus,Ida,Misao}.
On the other hand, because of the very nature of
the tidal force, a mass on the brane
exerts a force that tends to squeeze the matter
in the 5th direction at least in the far zone, meaning it works as an
attractive force. This means it is a negative energy
contribution, just like the Newtonian effective
gravitational energy.
It is therefore impossible to judge whether the
positivity of the ADM energy is guaranteed or not
within the present perturbation analysis.
Some global non-linear analysis will
be necessary to resolve this issue.

\section*{Acknowledgements}
We would like to thank J.~Garriga, G.~W.~Gibbons, H.~Kodama, A.~Ishibashi,
H.~Ishihara, T.~Tanaka and N.~Turok for discussions.
Part of this work was done while
MS and KM were participating the program, ``Structure Formation in the
Universe'', at the Newton Institute, University of Cambridge.
We are grateful to the Newton Institute for their hospitality.
MS's work is supported in part by Monbusho Grant-in-Aid
for Scientific Research No.~09640355.
TS's work is supported by JSPS Postdoctal Fellowship for
Research Abroad. KM's work is supported in part by Monbusho
Grant-in Aid for Specially Promoted Research No.~08102010.

\appendix

%
%
%
%

\section{Asymptotic Flatness at spatial infinity}

We briefly describe the definition of asymptotic
flatness and some useful results obtained in 4-dimensional
asymptotically flat spacetimes (see Refs.~\cite{AR,Gen}
for the details and exact descriptions ).
In this appendix the notation basically
follows Ref.~\cite{Gen}, which is slightly different from the
main text of this paper. Below the suffix $a,b,...$ denotes the
abstract index\cite{Wald}.

{\it Definition}: A spacetime ($M, q_{ab}$) will be
said to be asymptotically flat at spacelike infinity
${\cal I}$ if there exists a smooth function
$\Omega$ satisfying the following
features (i), (ii) and the energy momentum tensor satisfies the
fall off condition (iii);

\begin{list}{}{}
\item{(i)} $\Omega|_{{\cal I}}=0$ and $ d \Omega |_{{\cal I}} \neq 0$\,.

\item{(ii)}
 The following quantities have smooth limits on ${\cal I}$\,.
%
\begin{eqnarray}
r^a=\Omega^{-4}q^{ab}D_b \Omega\,,
\end{eqnarray}
%
%
\begin{eqnarray}
{\hat p}_{ab}=\Omega^2( q_{ab}
-F^{-1}\Omega^{-4}D_a \Omega D_b \Omega)=\Omega^2 p_{ab},
\end{eqnarray}
%
where $F=\mbox{\pounds}_r \Omega$\,.

\item{(iii)}
 $T _{\mu\nu}:=
 T_{ab}\base{e}{\mu}^a\base{e}{\nu}^b
= O( \Omega^4 )$ near ${\cal I}$,
where $\lbrace \base{e}{\mu}^a \rbrace_{\ssr \mu =0,1,2,3}$ is a tetrad of
the metric $q_{ab}$.
\end{list}

In this formalism, $\Omega \sim r^{-1}$.
For example, the extrinsic curvature of the $\Omega={\rm cosnt.}$
hypersurface is written as
%
\begin{eqnarray}
\kappa_{ab}:=\frac{1}{2}\mbox \pounds_{\hat r} p_{ab}
=\Omega^{-1} F^{1/2}{\hat p}_{ab}-\frac{1}{2}F^{-1/2}\mbox \pounds_r
{\hat p}_{ab},
\end{eqnarray}
%
where ${\hat r}^a:=r^a/{\sqrt {r^b r_b}} $. {}From the 4-dimensional
Einstein equations under the condition (i), (ii) and (iii),
we find
%
\begin{eqnarray}
F{\hat =}1~~~~{\rm and}~~~~{}^{(3)}{\hat R}_{ab}{\hat =}2{\hat p}_{ab},
\end{eqnarray}
%
where the hatted  equality, ${\hat =}$,
 denotes the evaluation on ${\cal I}$.
This gives the 3-dimensional Riemann tensor as
%
\begin{eqnarray}
{}^{(3)}{\hat R}_{abcd}{\hat =} 2{\hat p}_{a[c}{\hat p}_{d]b}\,,
\end{eqnarray}
%
which means that the 3-dimensional $\Omega=0$ surface is locally
a unit 3-dimensional timelike hyperboloid.

A part of the gravitational field is described by the electric part of
the 4-dimensional Weyl tensor,
%
\begin{eqnarray}
{}^{(4)}E_{ab}=\kappa_{ac}\kappa^c_b-\mbox \pounds_{\hat r} 
\kappa_{ab}+{\cal D}_{(a}a_{b)}-a_a a_b -\frac{1}{2}(p_a^c
p_b^d+p_{ab}{\hat r}^c {\hat r}^d)L_{cd},
\end{eqnarray}
%
where $a^a={\hat r}^b D_b {\hat  r}^a$ and
$L_{ab}={}^{(4)}R_{ab}-(1/6)q_{ab}{}^{(4)}R$. Around
$\Omega=0$, ${}^{(4)}E_{ab}$ is expanded as
${}^{(4)}E_{ab}=\sum_{\ell=0} {}^{(4)}E_{ab}^{(\ell)}\Omega^\ell$.
It can be shown that ${}^{(4)}E^{(0)}=0$ and
%
\begin{eqnarray}
{}^{(4)}E_{ab}^{(1)}=
\frac{1}{2}({\hat {\cal D}}_a {\hat {\cal D}}_{b}
+{\hat p}_{ab}) F^{(1)}-\frac{1}{2}(L_{ab}^{(3)}+p_{ab}
L_{cd}^{(3)}{\hat r}^c {\hat r}^d),
\end{eqnarray}
%
where $F=1+\sum_{\ell=1}F^{(\ell)}\Omega^\ell$.

{}From a rather lengthy argument,
 we find there exists an asymptotic conformal
killing vector, ${\hat \xi}^a$, such that
${\hat \xi}^a{\hat =}{\hat {\cal D}}^a \alpha$ and
$\mbox \pounds_{\hat \xi} {\hat p}_{ab}{\hat =}2\alpha {\hat p}_{ab}$
or ${\hat {\cal D}}_a {\hat {\cal D}}_b \alpha+{\hat p}_{ab}
\alpha{\hat =}0$. The conformal Killing vector induces the
time-translation symmetry at spatial infinity.
The asymptotic conformal Killing vector $t^\mu$ appeared
in Sec. IV is just ${\hat {\cal D}}^\mu \alpha$.

\end{document}